\documentclass[11pt,epsf,letterpaper]{article}%
\usepackage{setspace}
\usepackage{color}
\usepackage{amsmath}
\usepackage{amsfonts}
\usepackage{verbatim}
\usepackage{amssymb}
\usepackage{graphicx}%
\providecommand{\U}[1]{\protect\rule{.1in}{.1in}}
\begin{document}

\title{Generalized Hedgehog ansatz and Gribov copies in regions with non-trivial topologies}
\author{Fabrizio Canfora$^{1,2}$ Patricio Salgado-Rebolledo$^{1,3}$\\$^{1}$\textit{Centro de Estudios Cient\'{\i}ficos (CECS), Casilla 1469,
Valdivia, Chile.}\\$^{2}$\textit{Universidad Andr\'{e}s Bello, Av. Rep\'{u}blica 440, Santiago,
Chile.}\\$^{3}$\textit{Departamento de F\'{\i}sica, Universidad de Concepci\'{o}n,
Casilla 160-C, Concepci\'{o}n, Chile.}\\{\small canfora@cecs.cl, pasalgado@udec.cl}}
\maketitle

\begin{abstract}
In this paper the arising of Gribov copies both in Landau and Coulomb gauges
in regions with non-trivial topologies but flat metric, (such as closed tubes
$S^{1}\times D^{2}$, or $%
\mathbb{R}
\times T^{2}$) will be analyzed. Using a novel generalization of the hedgehog
ansatz beyond spherical symmetry, analytic examples of Gribov copies of the
vacuum will be constructed. Using such ansatz, we will also construct the
elliptic Gribov pendulum. The requirement of absence of Gribov copies of the
vacuum satisfying the strong boundary conditions implies geometrical
constraints on the shapes and sizes of the regions with non-trivial topologies.

\end{abstract}

\section{Introduction}

One of the most important sectors of the standard model is Yang-Mills theory
which describes QCD and the electro-weak theory. The degrees of freedom of a
gauge theory are encoded in the connection$\ (A_{\mu})^{a}$, which is a Lie
algebra valued one form. The action functional is invariant under gauge
transformations%
\begin{equation}
A_{\mu}\rightarrow U^{-1}A_{\mu}U+U^{-1}\partial_{\mu}U
\label{gaugetranformation}%
\end{equation}
whereas the physical observables are invariant under proper gauge
transformations. The latter has to be everywhere smooth and it has to decrease
at infinity in a suitable way (see, for instance, \cite{Benguria:1976in}).

Since it is still unknown how to use in practice gauge invariant variable in
Yang-Mills case\footnote{In the cases of topological field theories in 2+1
dimensions \cite{wittenjones} this goal has been partially achieved.}, the
most convenient practical choices for the gauge fixing are the Coulomb gauge
and the Landau gauge\footnote{The axial and light-cone gauge fixing choices
are affected by quite non-trivial problems (see, for instance, \cite{DeW03}%
).}. The standard approach to fix the gauge and to use Feynman expansion
around the trivial vacuum $A_{\mu}=0$ provides one with excellent results when
the coupling constant is small.

As it was for the first time discovered in \cite{Gri78}, a \textit{proper
gauge fixing} is not possible globally due to the presence of Gribov
copies\footnote{Furthermore, it has been shown by Singer \cite{singer}, that
if Gribov ambiguities occur in Coulomb gauge, they occur in all the gauge
fixing conditions involving derivatives of the gauge field (See also
\cite{jackiw0}). Other gauge fixings (such as the axial gauge, the temporal
gauge, and so on) free from gauge fixing ambiguities are possible but these
choices have their own problems (see, for instance, \cite{DeW03}).}. In the
QCD case, this effect is very important in the non-perturbative regime. Even
if one chooses a gauge free of Gribov copies, the effects of Gribov
ambiguities in other gauges generate a non-perturbative breaking of the BRST
symmetry \cite{Fuj}.

It has been suggested to exclude from the domain of the path-integral gauge
potentials $A_{\mu}$ which generate zero modes of the FP operator (see, in
particular, \cite{Gri78} \cite{Zw82} \cite{Zw89} \cite{DZ89}\ \cite{Zwa96}
\cite{Va92}; two nice reviews are \cite{SS05} \cite{EPZ04}). In this
framework, which is called the (refined) \textit{Gribov-Zwanziger approach} to
QCD, The region $\Omega$ around $A_{\mu}=0$ in which the FP operator is
positive is called Gribov region:%
\begin{align}
&  \Omega\overset{def}{=}\left\{  \left.  A_{\mu}\right\vert \ \partial^{\mu
}A_{\mu}=0\ \ and\ \ \det\partial^{\mu}D\left(  A\right)  _{\mu}>0\right\}
,\label{grireg}\\
D\left(  A\right)  _{\mu} &  =\partial_{\mu}+\left[  A_{\mu,\cdot}\right]
\ ,\label{covder}%
\end{align}
where $D\left(  A\right)  _{\mu}$ is the covariant derivative corresponding to
the field $A_{\mu}$. In the case in which the space-time metric is flat and
the \textit{topology is trivial} this approach coincides with usual
perturbation theory when the gauge field $A_{\mu}$ is close to the origin
(with respect to a suitable functional norm \cite{Va92}). At the same time,
this framework takes into account the infra-red effects related to the
partial\footnote{The condition to have a positive Faddeev-Popov operator is
not enough to completely eliminate Gribov copies in the Coulomb and Landau
gauges. It can be shown \cite{Va92} that there exist a smaller region (called
the \textit{modular region}) contained in the Gribov region which is free of
gauge fixing ambiguities. However, it is still not clear how to implement the
restriction to the modular region in practice.}\ elimination of the Gribov
copies \cite{Zw89} \cite{MaggS} \cite{Gracey}. When one takes into account the
presence of suitable condensates \cite{SoVae2} \cite{SoVar} \cite{SoVar2}
\cite{SoVar3} \cite{soreprl} the agreement with lattice data is
excellent\footnote{Recently, in \cite{criticism}, some criticism of this
approach has been proposed. However, this argument is only valid in the
perturbative framework (for instance, the gluons are assumed to be asymptotic
states) while the changes of the Gribov-Zwanziger approach are
non-perturbative in nature.} \cite{DOV} \cite{CucM}. Within the (refined)
\textit{Gribov-Zwanziger approach} to QCD, the \textit{copy-free neighborhood}
of the trivial vacuum $A_{\mu}=0$ \textit{has to be identified with}
\textit{perturbative region of the theory}. In this way, within the
perturbative region, the standard recipe to fix the gauge and perform the
perturbative expansion makes sense thanks to the lack of any overcounting. The
appearance of copies of the vacuum satisfying the strong boundary conditions
is one of the worst pathologies of perturbation theory.

On the other hand, it is a well established fact by now that Yang-Mills theory
may have knotted excitations \cite{knotted} \cite{knotted1} \cite{knotted2}
(the simplest non-trivial example corresponding to a closed tube: the "donut"
or \textit{unknot} with topology $S^{1}\times D^{2}$). Moreover, as it is well
known (see, for instance, \cite{infinitevolume}), non-trivial topologies such
as $%
\mathbb{R}
\times T^{2}$ and $%
\mathbb{R}
\times T^{3}$ are very important to understand the infinite volume limit of
Yang-Mills theory as well as chiral symmetry breaking generated by the Casimir
force (a related reference is \cite{casimir1}). In order to describe these
situations one has to be able to define consistently Yang-Mills theory inside
bounded regions with the non-trivial topologies.

The main goal of the present paper is to analyze whether or not Gribov copies
of the vacuum can actually exist inside a space-time region of the topology
such as $S^{1}\times D^{2}$ (but more general topologies relevant, for
instance, for the infinite volume limit of Yang-Mills as well as lattice
QCD-such as $T^{3}$ and $%
\mathbb{R}
\times T^{2}$, $%
\mathbb{R}
\times T^{3}$ as well as the interior of an ellipse-will be considered). The
absence of Gribov copies of the vacuum (satisfying strong boundary conditions)
is a necessary condition for the existence of a perturbative region in the
functional space around $A_{\mu}=0$. It will be shown that such requirement
implies non-trivial restrictions on the possible sizes and shapes of the
corresponding regions. In a sense, these results are more surprising than the
ones obtained in \cite{CGO} \cite{ACGO} \cite{CGO2} in which it has been shown
that the pattern of appearance of Gribov copies strongly depends on the
space-time metric. All the examples of non-trivial copies of the vacuum will
be in flat metrics (but with non-trivial topologies). The present analysis
will also show that the Gribov phenomenon strongly depends on the shapes and
sizes of the bounded regions where one wants to study gauge theories.

Unlike the equation for the zero-modes of the Faddeev-Popov operator, the
issue of the appearance of Gribov copies of the vacuum is non-linear in nature
and therefore can be quite complicated when non-standard topologies are
considered. The technical tool necessary in order to analyze such an issue is
a novel (to the best of authors knowledge) generalization of the
\textit{hedgehog ansatz} beyond spherical symmetry for the non-linear
sigma-model. Such a generalization is quite interesting in itself since it
allows to reduce the non-linear system of coupled partial differential
equations corresponding to the equations of motion of the non-linear sigma
model to a single non-linear partial differential equation which can be
analyzed with the tools of solitons theory.

The paper is organized as follows. In section two the notion of strong
boundary conditions with non-trivial topologies will be analyzed. In section
three, the relations between the Gribov copies equation and the non-linear
sigma-model will be discussed. In section four, the generalized hedgehog
ansatz will be constructed. In sections five and six various non-trivial
examples of copies of the vacuum will be described. Some conclusions will be
drawn in the last section. In the Appendix a novel way to\ implement both
spherical and elliptical symmetries (which could be useful in the context of
calorons) will be presented.

\section{Strong boundary conditions with non-trivial topologies}

Here, we will first review the definition of strong boundary conditions in the
case in which the metric is flat and the topology trivial. In the present
paper we will mainly analyze the $SU(2)$ case but many of the present results
also extend to other Lie groups. A useful starting point is the definition of
non-Abelian charges $Q^{a}$ (see \cite{Benguria:1976in} \cite{Gri78}
\cite{abbdeser}, \cite{jackiw}; see, for a detailed review, \cite{lavelle}):%
\begin{equation}
Q^{\left(  a\right)  }=\int_{M}d^{3}x\partial_{i}E^{ia}=\lim_{R\rightarrow
\infty}\int_{\Sigma_{R}}d^{2}\Sigma\left(  n_{i}E^{ia}\right)  \ ,
\label{charge1}%
\end{equation}
where $M$ is the constant-time hypersurface in four-dimensional Minkowski
space, $\Sigma_{R}$ is the two-dimensional sphere of radius $R$, $n_{i}$ is
the outer pointing unit normal to $\Sigma_{R}$, the indices $i$, $j$, $k$
refers to space-like directions while $a$, $b$, $c$,...are internal $SU(2)$ indices.

Thus, one can define a \textit{proper gauge transformation} $U$ as a smooth
gauge transformation which does not change the value of the charge as a
surface integral at infinity:%
\begin{equation}
\lim_{R\rightarrow\infty}\int_{\Sigma_{R}}d^{2}\Sigma\left(  n_{i}%
E^{ia}\right)  =\lim_{R\rightarrow\infty}\int_{\Sigma_{R}}d^{2}\Sigma\left(
n_{i}\left(  U^{-1}EU\right)  ^{ia}\right)  \ . \label{proper1}%
\end{equation}
Therefore, $U$ is proper if it is smooth and approaches to an element of the
center of the gauge group at spatial infinity.

A Gribov copy $U$ on a flat and topologically trivial space-time
satisfies\textit{ the strong boundary} conditions if $U$ is proper. A copy of
this type is particularly problematic since it would represent a failure of
the whole gauge fixing procedure. Indeed, if the vacuum $A_{\mu}=0$ possesses
a copy fulfilling the strong boundary conditions, not even usual perturbation
theory leading to the standard Feynman rules in the Landau or Coulomb gauge
would be well defined.

\subsection{Strong boundary conditions on $S^{1}\times D^{2}$}

The simplest topology corresponding to a closed knotted tube is $\Upsilon
=S^{1}\times D^{2}$: this case is relevant in the analysis of non-trivial
topological excitations of glue-balls. In particular, there is a sound
evidence supporting the existence of excited glue-ball states with the
topology of $\Upsilon=S^{1}\times D^{2}$ (see, in particular, \cite{knotted1}
\cite{knotted2}). In this case, the spatial metric describing the region
$\Upsilon$ reads:%
\begin{align}
ds^{2}  &  =\left(  d\phi\right)  ^{2}+\left(  dr^{2}+r^{2}d\theta^{2}\right)
\ ,\label{metr1}\\
0  &  \leq\phi\leq2\pi\ ,\ \ 0\leq\theta\leq2\pi\ ,\ \ 0\leq r\leq
R\ ,\nonumber
\end{align}
where the coordinate $\phi$\ corresponds to the $S^{1}$ circle of $\Upsilon$,
while the coordinates $r$ and $\theta$ describe the disk $D^{2}$ of $\Upsilon$.

The radius of the disk is $R$ and the boundary of $\Upsilon$ is:%
\[
\partial\Upsilon=S^{1}\times\partial D^{2}\ .
\]
Thus, in this case a smooth gauge transformation $U$ is proper if%
\[
\left.  U\right\vert _{S^{1}\times\partial D^{2}}\in Z_{2}\ ,
\]
where $Z_{2}$ is the center\footnote{Obviously, the center of $SU(2)$ is made
of two elements: $\pm\mathbf{1}$.} of $SU(2)$. In other words, a smooth gauge
transformation $U=U(\phi,r,\theta)$ is proper if%
\begin{align}
U(\phi+2m\pi,r,\theta+2n\pi)  &  =U(\phi,r,\theta)\ ,\ \ m,n\in%
\mathbb{Z}
\ ,\label{tub1}\\
U(\phi,R,\theta)  &  \in Z_{2}\ ,\ \forall\ \phi,\ \theta\ . \label{tub2}%
\end{align}

\subsection{strong boundary conditions on $T^{3}$}

In the cases in which the spatial topology is $T^{3}$ (which, for instance, is
relevant in the case of lattice QCD), the flat spatial metric describing
$T^{3}$ reads:%
\begin{align}
ds^{2}  &  =\sum_{i=1}^{3}\lambda_{i}^{2}\left(  d\phi_{i}\right)
^{2}\ ,\ \ \lambda_{i}\in%
\mathbb{R}
\ ,\label{metr2}\\
0  &  \leq\phi_{i}\leq2\pi\ ,\nonumber
\end{align}
where the coordinate $\phi_{i}$\ corresponds to the $i$-th factor $S^{1}$ in
$T^{3}$ while $\lambda_{i}$ represents the size of the $i$-th factor $S^{1}$
in $T^{3}$. In this case, due to the fact that $\partial T^{3}=0$, a smooth
gauge transformation $U=U(\phi_{1},\phi_{2},\phi_{3})$ is proper if%

\begin{equation}
U(\phi_{1}+2m_{1}\pi,\phi_{2}+2m_{2}\pi,\phi_{3}+2m_{3}\pi)=U(\phi_{1}%
,\phi_{2},\phi_{3})\ ,\ \ m_{i}\in%
\mathbb{Z}
\ . \label{tub3}%
\end{equation}

\subsection{strong boundary conditions on $%
\mathbb{R}
\times T^{2}$}

The cases in which the spatial topology is $%
\mathbb{R}
\times T^{2}$ may correspond to situations in which the bounded region in
which we want to define perturbative Yang-Mills theory is much longer in one
direction (the $x$ axis) with respect to the other two.

It is worth emphasizing here that, in the Euclidean theory, the topology $%
\mathbb{R}
\times T^{3}$ is also very important in order to understand both the infinite
volume limit in QCD (see, for instance, \cite{infinitevolume}) and effects of
chiral symmetry breaking in QCD (see, for instance, \cite{casimir1}). In
particular, it has been shown in \cite{casimir1} the relevance of the
Casimir\footnote{Indeed, Casimir force is a genuine finite-size effect which,
obviously, is not visible if the theory is analyzed in unbounded regions.}
force to understand chiral symmetry breaking. Hence, the importance of
finite-size effects to explain these phenomena and the necessity to have a
well-defined perturbation theory in such bounded regions with non-trivial
topology require a deeper understanding of the Gribov problem in bounded
region as well. As it will be shown in the following, the present results on
the Coulomb gauge can be trivially extended to the case of the Landau gauge
defined on $%
\mathbb{R}
\times T^{3}$. Thus, the gauge-fixing pathologies arising in this case have to
be carefully taken into account.

The flat spatial metric describing $%
\mathbb{R}
\times T^{2}$ reads:%
\begin{align}
ds^{2}  &  =dx^{2}+\sum_{i=1}^{2}\lambda_{i}^{2}\left(  d\phi_{i}\right)
^{2}\ ,\ \ \lambda_{i}\in%
\mathbb{R}
\ ,\label{metr2.5}\\
0  &  \leq\phi_{i}\leq2\pi\ ,\nonumber
\end{align}
where the coordinate $\phi_{i}$\ corresponds to the $i$-th factor $S^{1}$ in
$T^{2}$ while $\lambda_{i}$ represents the size of the $i$-th factor $S^{1}$
in $T^{2}$. In this case, $\partial\left(
\mathbb{R}
\times T^{2}\right)  $ is non-trivial and can be identified with the two limit
point $x\rightarrow\pm\infty$ of $%
\mathbb{R}
$. Thus, a smooth gauge transformation $U=U(\phi_{1},\phi_{2},x)$ is proper if%

\begin{align}
U(\phi_{1}+2m_{1}\pi,\phi_{2}+2m_{2}\pi,x)  &  =U(\phi_{1},\phi_{2}%
,x)\ ,\ \ m_{i}\in%
\mathbb{Z}
\ ,\label{tub3.5}\\
\underset{x\rightarrow\pm\infty}{\lim}U(\phi_{1},\phi_{2},x)  &  \in
Z_{2}\ ,\ \forall\ \phi_{1},\ \phi_{2}\ . \label{tub3.75}%
\end{align}

\section{Gribov copies and non-linear sigma model}


In the present paper we will analyze the copies of the vacuum in the Coulomb
gauge but many of the present results can be easily extended to the Landau
gauge. Since the Gribov copies of the vacuum in the Coulomb gauge can be seen
as the Euler-Lagrange equations corresponding to a non-linear sigma model, we
will first review some basic features of this model. The non-linear sigma
model Lagrangian in $D$ space-like dimensions can be written in terms of group
valued scalar field $U$. In the following we will consider the $SU(2)$ group,
so that the corresponding Lagrangian is:%
\begin{align}
S  &  =\frac{\kappa^{2}}{2}\int\sqrt{-g}d^{D}xTr\left[  R^{i}R_{i}\right]
\ ,\label{symh}\\
R_{i}  &  =U^{-1}\partial_{i}U\ , \label{skyrme2}%
\end{align}%
\begin{align*}
R_{i}  &  =R_{i}^{a}t_{a}\ ,\\
t^{a}t^{b}  &  =-\delta^{ab}\mathbf{1}-\varepsilon^{abc}t^{c}\ ,\ \\
\left[  R_{i},R_{j}\right]  ^{c}  &  =C_{ab}^{c}R_{i}^{a}R_{j}^{b}%
\ ,\ C_{ab}^{c}=-2\varepsilon_{cab}\ ,\\
\varepsilon^{abc}\varepsilon^{mnc}  &  =\left(  \delta_{a}^{m}\delta_{b}%
^{n}-\delta_{a}^{n}\delta_{b}^{m}\right)  \ ,
\end{align*}
where $D$ is the number of space dimensions, $\kappa$ is the sigma-model
coupling constants, $g$ is the determinant of the space-like metric,
$\mathbf{1}$ is the identity $2\times2$ matrix and $t^{a}$ the generator of
$SU(2)$ (where the Latin letters $a$, $b$, $c$ and so on corresponds to group
indices). The equation characterizing the appearance of Gribov copies of the
vacuum in the Coulomb gauge (which corresponds to the Euler-Lagrange equations
of the action in Eq. (\ref{symh})) read:%
\begin{equation}
\nabla^{i}R_{i}=\nabla^{i}\left(  U^{-1}\partial_{i}U\right)  =0\ ,
\label{nonlinearsigma1}%
\end{equation}
where $\nabla^{i}$ is the Levi-Civita covariant derivative corresponding to
the metric (which, in the present case will have one of the forms in Eqs.
(\ref{metr1}), (\ref{metr2}) and (\ref{metr2.5})). The case of the Landau
gauge is analogous.

The following standard parametrization of the $SU(2)$-valued functions
$U(x^{i})$ is useful:%
\begin{align}
U(x^{i})  &  =Y^{0}\mathbf{1}+Y^{a}t_{a}\ ,\ U^{-1}(x^{i})=Y^{0}%
\mathbf{1}-Y^{a}t_{a}\ ,\label{standard1}\\
Y^{0}  &  =Y^{0}(x^{i})\ ,\ \ \ Y^{a}=Y^{a}(x^{i})\ ,\label{standard2}\\
\left(  Y^{0}\right)  ^{2}+Y^{a}Y_{a}  &  =1\ , \label{standard3}%
\end{align}
where, of course, the sum over repeated indices is understood also in the case
of the group indices (in which case the indices are raised and lowered with
the flat metric $\delta_{ab}$). Therefore, the $R_{i}$ in Eq. (\ref{skyrme2})
can be written as follows:%
\begin{equation}
R_{i}^{c}=\varepsilon^{abc}Y_{a}\partial_{i}Y_{b}+Y^{0}\partial_{i}Y^{c}%
-Y^{c}\partial_{i}Y^{0}\ . \label{standard4}%
\end{equation}

\section{Generalized Hedgehog ansatz}

Due to the intrinsic non-linear nature of the Gribov copies of the vacuum, it
is necessary to introduce a suitable technical tool which allows to study such
phenomenon with non-trivial topologies.

Here we will first discuss the geometrical interpretation of \textit{hedgehog
ansatz} as an effective tool to reduce the field equations of the non-linear
sigma model (which, generically, are a \textit{system of coupled non-linear
partial differential equations}) to a single scalar non-linear partial
differential equation. The geometrical analysis of such very important feature
of the hedgehog ansatz allows to construct the natural generalization of the
hedgehog ansatz for the non-linear sigma model. This section is interesting in
itself since, to the best of authors knowledge, this is the first systematic
reduction of the equations of motion of the non-linear sigma model (which is a
system of coupled non-linear partial differential equations) to a single
non-linear scalar equation beyond spherical symmetry. In terms of the group
element $U$ the standard spherically symmetric hedgehog ansatz reads%
\begin{align}
U  &  =\mathbf{1}\cos\alpha(r)+\widehat{n}^{a}t_{a}\sin\alpha(r)\ ,\ \ U^{-1}%
=\mathbf{1}\cos\alpha(r)-\widehat{n}^{a}t_{a}\sin\alpha
(r)\ ,\label{hedgehog1.1}\\
\ \widehat{n}^{1}  &  =\sin\theta\cos\phi\ ,\ \ \ \widehat{n}^{2}=\sin
\theta\sin\phi\ ,\ \ \ \widehat{n}^{3}=\cos\theta\ , \label{hedgehog1.111}%
\end{align}
where $r$, $\theta$ and $\phi$ are spherical coordinates of the flat Euclidean
metric, the $\widehat{n}^{a}$ are normalized with respect to the internal
metric $\delta_{ab}$. In terms of the variables $Y^{0}$ and $Y^{a}$ the
hedgehog ansatz corresponds to the following choice:%
\begin{equation}
Y^{0}=\cos\alpha(r)\ ,\ \ Y^{a}=\widehat{n}^{a}\sin\alpha(r)\ ,\ \ \delta
_{ab}\widehat{n}^{a}\widehat{n}^{b}=1\ . \label{hedgehog1.2}%
\end{equation}
Thus, the expression for $R_{i}^{a}$ in Eq. (\ref{standard4}) reads%
\begin{equation}
R_{i}^{c}=\delta^{dc}\left(  \sin^{2}\alpha\right)  \varepsilon_{abd}%
\widehat{n}^{a}\partial_{i}\widehat{n}^{b}+\widehat{n}^{c}\partial_{i}%
\alpha+\frac{\sin\left(  2\alpha\right)  }{2}\partial_{i}\widehat{n}^{c}\ .
\label{hedgehog1.5}%
\end{equation}
The equations of motion for the non-linear sigma model in Eq.
(\ref{nonlinearsigma1}) corresponding to this ansatz read%
\begin{equation}
\left(  \nabla^{i}\partial_{i}\alpha\right)  +L\sin2\alpha=0\ ,
\label{sinegordon1}%
\end{equation}
where $L$ is a suitable function of the radial coordinate $r$ (see below).
Indeed, this is a very important and non-trivial characteristic of the
spherically symmetric hedgehog ansatz which reduces a system of coupled
non-linear partial differential equations to a single scalar equation. It is
easy to see that in the case of the flat three-dimensional Euclidean metric on
$%
\mathbb{R}
^{3}$ in spherical coordinates the above reduces to the standard Gribov
pendulum of the vacuum (see \cite{Gri78} \cite{SS05}).

The previous analysis suggests to analyze which are the geometrical conditions
which allow to reduce a system of non-linear coupled Partial Differential
Equations to a single scalar PDE. Let us consider the following generalization
of the hedgehog ansatz:%
\begin{align}
U  &  =\mathbf{1}\cos\alpha+\widehat{n}^{a}t_{a}\sin\alpha\ ,\ \ U^{-1}%
=\mathbf{1}\cos\alpha-\widehat{n}^{a}t_{a}\sin\alpha\ ,\label{hedgehog1.75}\\
\delta_{ab}\widehat{n}^{a}\widehat{n}^{b}  &  =1\ ,\ \ \alpha=\alpha\left(
x^{i}\right)  \ ,\ \ \widehat{n}^{a}=\widehat{n}^{a}\left(  x^{i}\right)  \ ,
\label{hedgehog1.95}%
\end{align}
such that%
\begin{align}
\left(  \partial_{i}\alpha\right)  \left(  \nabla^{i}\widehat{n}^{a}\right)
&  =0\ ,\label{hedgehogcondition3}\\
\left(  \nabla^{i}\partial_{i}\right)  \widehat{n}^{c}  &  =2L\widehat{n}%
^{c}\ , \label{hedgehogcondition4}%
\end{align}
where $L$ (which has to be the same for all the non-vanishing\footnote{Note
that, in order for the solution to be non-trivial, at least two of the three
$\widehat{n}^{c}$\ have to be non-vanishing.} $\widehat{n}^{c}$) may depend on
the space-time coordinates.

The important point is that, when the conditions (\ref{hedgehogcondition3})
and (\ref{hedgehogcondition4})\ are satisfied, the function $\alpha$ (which in
the usual spherically symmetric hedgehog ansatz can depend only on the radial
coordinate $r$) can now depend on any set of coordinates and the functions
$\widehat{n}^{a}$ (which in the usual case have to coincide with the unit
radial vector) can be adapted to cases in which there is no spherical symmetry
(and, therefore, no natural radial coordinate). Indeed, the expression for
$R_{i}^{a}$ in Eq. (\ref{standard4}) reads%
\begin{equation}
R_{i}^{c}=\delta^{dc}\left(  \sin^{2}\alpha\right)  \varepsilon_{abd}%
\widehat{n}^{a}\partial_{i}\widehat{n}^{b}+\widehat{n}^{c}\partial_{i}%
\alpha+\frac{\sin\left(  2\alpha\right)  }{2}\partial_{i}\widehat{n}^{c}\ .
\label{hedgehog2}%
\end{equation}
Hence, provided Eqs. (\ref{hedgehogcondition3}) and (\ref{hedgehogcondition4})
are fulfilled and taking into account that the expression%
\[
\left(  \nabla^{i}\widehat{n}^{a}\right)  \left(  \partial_{i}\widehat{n}%
^{b}\right)  =g^{ij}\left(  \partial_{i}\widehat{n}^{a}\right)  \left(
\partial_{j}\widehat{n}^{b}\right)
\]
is symmetric under the exchange of $a$ and $b$ (so that its contraction with
$\varepsilon_{abd}$\ vanishes) the coupled system of non-linear Partial
Differential Equations (PDE) corresponding to the equations of motion of the
non-linear sigma model reduces to the single scalar non-linear partial
differential equation of sine-Gordon type:%
\begin{equation}
\left(  \nabla^{i}\partial_{i}\alpha\right)  +L\sin2\alpha=0\ .
\label{sinegordon1.5}%
\end{equation}

Indeed, it is easy to see that in the usual spherically symmetric flat ansatz
in Eqs. (\ref{hedgehog1.1}) and (\ref{hedgehog1.111}) satisfies the conditions
in Eqs. (\ref{hedgehogcondition3}) and (\ref{hedgehogcondition4}) with
$L=-1/r^{2}$ and that, correspondingly, Eq. (\ref{sinegordon1.5}) reduces to
the usual Gribov pendulum (see \cite{Gri78} \cite{SS05}). However, it is worth
noting here that the present derivation is far simpler (just three lines of
computation) and, furthermore, it discloses in a very clear way the geometry
behind the hedgehog ansatz.

The usual Gribov pendulum equation on flat and topologically trivial
three-dimensional spaces (see \cite{Gri78} \cite{SS05}) does not coincide with
the sine-Gordon equation due to the radial dependence of the determinant of
the three-dimensional Euclidean metric in spherical coordinates. For these
reasons, there are no analytic examples of copies of the vacuum in this case.
On the other hand, using the present generalized hedgehog ansatz, one can
construct many analytic examples of copies of the vacuum. These examples are
very useful since they allow to express the requirement of absence of copies
of the vacuum in terms of explicit constraints on the shapes and sizes of the
corresponding bounded regions where one wants to define a perturbative region.

\subsection{Regularity conditions}

Another important requirement to satisfy the strong boundary conditions is the
regularity of the pure gauge field $U^{-1}dU$ everywhere. In the standard
spherically-symmetric case, the only problematic point is the origin of the
coordinates in which the spherical coordinates system is not well defined. One
way to understand this condition in general is to analyze the behavior of the
one-form
\begin{equation}
R^{c}=R_{i}^{c}\wedge dx^{i} \label{regularity1}%
\end{equation}
(where $R_{i}^{c}$ is given by (\ref{hedgehog2}) and $\alpha$ satisfies
(\ref{sinegordon1.5})) when one changes coordinates from the Cartesian to a
coordinates system which is singular somewhere. Indeed, in the following we
will need to use non-Cartesian coordinates systems which are adapted to the
symmetry of the problem (such as the elliptic coordinates systems in the next sections).

In the case of Cartesian coordinates, the regularity at the origin can be read
by looking at the components of the one-form $R_{i}^{c}$ since the coordinates
system is regular everywhere. On the other hand, if one is in a spherical or
elliptical coordinates system, the angular coordinates are not well defined at
the origin and some extra care is required. For instance, to disclose the
singularity of the one forms $d\theta$ and $d\phi$ at the origin one can
analyze the Jacobian of the transformation from Cartesian coordinates
($x^{\prime i}$) to spherical coordinates ($x^{j}$):%
\begin{equation}
dx^{\prime i}=J_{j}^{i}dx^{j}\ . \label{regularity2}%
\end{equation}
Due to the regularity of the Cartesian coordinates, the Jacobian itself
encodes the informations of the singularity at the origin. In the spherically
symmetric case, for example, if one takes into account the Jacobian, the
regularity condition at the origin coincides\footnote{In the spherical case
the factor $\sin\theta$ which appears in the denominator of many components of
the Jacobian is canceled out by a similar factor in the numerator of the
spherically symmetric hedgehog ansatz in the normalized internal vector
$\widehat{n}^{i}$.} with the usual one derived in Cartesian coordinates
\cite{SS05}.

The requirement for regularity can be described as follows. Let's call $x_{s}$
a point in which the coordinates are singular and let's consider the following
behavior for $\alpha$ when $x\rightarrow x_{s}$
\begin{equation}
\alpha\left(  x\right)  \underset{x\rightarrow x_{s}}{\rightarrow}n\pi+\beta
f\left(  x\right)  \label{regularity3}%
\end{equation}
with $f\left(  x\right)  \underset{x\rightarrow x_{s}}{\rightarrow}0$ and
$\beta$ an arbitrary constant. Then, a copy of the form (\ref{hedgehog2})
satisfy
\begin{equation}
R_{i}^{c}\underset{x\rightarrow x_{s}}{\rightarrow}\beta^{2}\delta^{dc}%
f^{2}\varepsilon_{abd}\widehat{n}^{a}\partial_{i}\widehat{n}^{b}+\beta
\widehat{n}^{c}\partial_{i}f+\beta f\partial_{i}\widehat{n}^{c}\ .
\label{regularity4}%
\end{equation}
Then using (\ref{regularity2}) we see that $f(x)$ must be such that%
\begin{equation}
R_{i}^{c}\left(  J^{-1}\right)  _{j}^{i} \label{regularity5}%
\end{equation}
(where $\left(  J^{-1}\right)  _{j}^{i}$ is the inverse Jacobian of the
transformation from Cartesian to the curvilinear coordinates system of
interest) are regular functions in $x=x_{s}$.

\section{Explicit examples of vacuum copies in non-trivial topologies}

Here we will discuss many example of vacuum copies.

\subsection{$S^{1}\times D^{2}$ topology}

Here we will consider the following ansatz corresponding to the metric in Eq.
(\ref{metr1}):
\[
\alpha=\alpha(r)\ ,\ \ \widehat{n}^{1}=\cos\left(  \theta\right)
\ ,\ \ \widehat{n}^{2}=\sin\left(  \theta\right)  \ ,\ \widehat{n}^{3}=0\ .
\]
It is easy to see that the above ansatz satisfies the conditions in Eqs.
(\ref{hedgehogcondition3}) and (\ref{hedgehogcondition4}) with the following
$L$:%
\begin{align*}
\left(  \nabla^{i}\partial_{i}\right)  \widehat{n}^{1}  &  =-\frac{1}{r^{2}%
}\widehat{n}^{1}\ ,\ \ \left(  \nabla^{i}\partial_{i}\right)  \widehat{n}%
^{2}=-\frac{1}{r^{2}}\widehat{n}^{2}\ \Rightarrow\\
L  &  =-\frac{1}{2r^{2}}\ .
\end{align*}
Thus, in this case the equation (\ref{sinegordon1.5}) for the copy of the
vacuum reduces to%
\begin{equation}
r\partial_{r}\left(  r\partial_{r}\alpha\right)  -\gamma\sin2\alpha
=0\ \Leftrightarrow\label{caso1.01}%
\end{equation}%
\begin{align}
\frac{\partial^{2}\alpha}{\partial\tau^{2}}  &  =\gamma\sin2\alpha
\ ,\ \ \gamma=\frac{1}{2}\ ,\ \ \tau-\tau_{0}=\log r\ ,\label{caso1.1}\\
E  &  =\frac{1}{2}\left[  \left(  \frac{\partial\alpha}{\partial\tau}\right)
^{2}+\gamma\cos2\alpha\right]  \ ,\label{caso1.2}\\
\tau-\tau_{0}  &  =\pm\int_{\alpha(\tau_{0})}^{\alpha(\tau)}\frac{dy}%
{\sqrt{2E-\gamma\cos2y}}\ , \label{caso1.3}%
\end{align}
where $\tau_{0}$ and $E$ are integration constants.

In terms of the radial coordinate $r$ one has to require that the copy is
regular at the origin and that it approaches an element of the center when
$r=R$ which is the boundary of $D^{2}$. Without loss of generality one can
require (see Eq. (\ref{hedgehog1.75})):%
\begin{equation}
\alpha\underset{r\rightarrow0}{\rightarrow}0\ ,\ \ \alpha(R)=n\pi\ ,\ \ \ n\in%
\mathbb{Z}
\ . \label{condition1}%
\end{equation}
In terms of the variable $\tau$ the above conditions read%
\begin{align}
&  \alpha\underset{\left(  \tau-\tau_{0}\right)  \rightarrow-\infty
}{\rightarrow}0\ ,\ \ \alpha\underset{\left(  \tau-\tau_{0}\right)
\rightarrow\tau^{\ast}}{\rightarrow}n\pi\ ,\label{condition2}\\
\tau^{\ast}  &  =\log R\ .\nonumber
\end{align}
Hence, we have to analyze under which conditions on the parameter $\gamma$ and
on the integration constants $E$ and $\tau_{0}$ Eq. (\ref{caso1.2}) can have
solutions satisfying the conditions in Eq. (\ref{condition2}). The simplest
way to answer to this question is to interpret Eq. (\ref{caso1.2}) as the
energy conservation of the following one-dimensional problem in which $\tau$
plays the role of the effective time :%
\begin{align}
\overline{E}  &  =\frac{1}{2}\left(  \frac{\partial A}{\partial\tau}\right)
^{2}+V\left(  A\right)  \ ,\label{1Deffective1}\\
V(A)  &  =2\gamma\cos A\ ,\ \ \ A=2\alpha\ ,\ \ \ \overline{E}=4E\ .
\label{1Deffective2}%
\end{align}
The boundary conditions in Eq. (\ref{condition2}) in terms of $A$ read (we
will consider the case $n=1$)%
\begin{equation}
A\underset{\left(  \tau-\tau_{0}\right)  \rightarrow-\infty}{\rightarrow
}0\ ,\ \ A\underset{\left(  \tau-\tau_{0}\right)  \rightarrow\tau^{\ast}%
}{\rightarrow}2\pi\ . \label{1Deffectiveboundcond1}%
\end{equation}
The above conditions mean that $0$ and $2\pi$ have to be two turning points
and that the corresponding period has to diverge:%
\begin{equation}
\frac{1}{\sqrt{2}}\int_{0}^{2\pi}\frac{dy}{\sqrt{\overline{E}-2\gamma\cos y}%
}\rightarrow\infty\ . \label{caso1.4}%
\end{equation}
In order to satisfy this condition, it is enough to choose%
\begin{equation}
\overline{E}=2\gamma=1\ , \label{csao1.5}%
\end{equation}
the explicit form for the copy being%
\begin{equation}
\alpha=2\arctan\left[  \frac{r}{\bar{r}_{0}}\right]  \label{spaccspacc}%
\end{equation}
with $\bar{r}_{0}$ an integration constant (whose role will be described in a moment).

In terms of the original radial variable $r$, the condition in Eq.
(\ref{condition1}) can only be satisfied\footnote{The reason is that the copy
profile in Eq. (\ref{spaccspacc}) is an increasing function from $0$ to
$\infty$ and its maximum is $\pi$. Note that, however, the copy profile
approaches the asymptotic value very rapidly.} if $R\rightarrow\infty$. In
physical terms, this means that the condition in Eq. (\ref{condition1}) can be
fulfilled when $R$ is very large compared to $\bar{r}_{0}$:
\[
R\gg\bar{r}_{0}.
\]
Obviously, in the present case the only two natural lengths are the radius of
the disk $R$ and the perimeter of the $S^{1}$ factor (which has been set to
$2\pi$) of the donut $S^{1}\times D^{2}$ and consequently $\bar{r}_{0}$
represents the size of $S^{1}$. Thus, the previous analysis tells that when
the radius of the disk is much larger than the perimeter of $S^{1}$ the donut
is on the verge of supporting smooth copies of the vacuum satisfying the
strong boundary conditions. Hence, in order to avoid this pathology the donut
cannot be too "fat".

It is easy to check that such copy is regular at the origin. The components of
$U^{-1}dU$ in Cartesian coordinates for this case are given by%
\[
R_{i}^{c}\left(  J^{-1}\right)  _{j}^{i}\text{ \ },\text{ \ }i=r,\theta\text{
\ },\text{ \ }j=1,2
\]
With%
\[
J^{-1}=\left(
\begin{array}
[c]{cc}%
\cos(\theta) & \sin(\theta)\\
-\frac{\sin(\theta)}{r} & \frac{\cos(\theta)}{r}%
\end{array}
\right)
\]
For $r\rightarrow0$ copies must decay as%
\begin{equation}
\alpha\underset{r\rightarrow0}{\rightarrow}\beta r\text{ }, \label{ec}%
\end{equation}
for some $\beta\in%
\mathbb{R}
$ and the profile in Eq. (\ref{spaccspacc}) satisfies this condition.

\subsection{$T^{3}$ topology}

As it has been already emphasized, the $T^{3}$ topology is relevant, for
instance, in the analysis of the thermodynamical limit. Here we will consider
the following ansatz corresponding to the metric in Eq. (\ref{metr2}):
\begin{align}
\alpha &  =\alpha(\phi_{1})\ ,\ \ \widehat{n}^{1}=\cos\left(  p\phi_{2}%
+q\phi_{3}\right)  \ ,\ \ \widehat{n}^{2}=\sin\left(  p\phi_{2}+q\phi
_{3}\right)  \ ,\label{caso2}\\
\widehat{n}^{3}  &  =0\ ,\ \ p,q\in%
\mathbb{Z}
\ .\nonumber
\end{align}
It is easy to see that the above ansatz satisfies the conditions in Eqs.
(\ref{hedgehogcondition3}) and (\ref{hedgehogcondition4}) with the following
$L$:%
\[
L=-\frac{1}{2}\left(  \left(  \frac{p}{\lambda_{2}}\right)  ^{2}+\left(
\frac{q}{\lambda_{3}}\right)  ^{2}\right)  \ .
\]
Thus, in this case the equation (\ref{sinegordon1.5}) for the copy of the
vacuum reduces to a flat elliptic sine-Gordon equation:%
\begin{align}
\left(  \frac{\partial^{2}}{\partial\phi_{1}^{2}}\right)  \alpha &
=\gamma\sin2\alpha\ ,\label{caso2.1}\\
\gamma &  =\frac{\lambda_{1}^{2}}{2}\left(  \left(  \frac{p}{\lambda_{2}%
}\right)  ^{2}+\left(  \frac{q}{\lambda_{3}}\right)  ^{2}\right)  \ .
\label{caso2.11}%
\end{align}
In order for a solution of Eq. (\ref{caso2.1}) to define a copy of the vacuum
satisfying the strong boundary condition (see Eqs. (\ref{hedgehog1.75}) and
(\ref{tub3})) it is necessary to require%
\begin{equation}
\alpha(\phi_{1}+2m\pi)=\alpha(\phi_{1})\ ,\ \ m\in%
\mathbb{Z}
\ , \label{caso2.2}%
\end{equation}
where we will consider $m=1$ in the following. As in the previous subsection,
Eq. (\ref{caso2.1}) can be reduced to a first order conservation law:%
\begin{align}
E  &  =\frac{1}{2}\left[  \left(  \frac{\partial\alpha}{\partial\phi_{1}%
}\right)  ^{2}+\gamma\cos2\alpha\right]  \ ,\label{caso2.22}\\
\phi_{1}-\phi_{0}  &  =\pm\int_{\alpha(\phi_{0})}^{\alpha(\phi_{1})}\frac
{dy}{\sqrt{2E-\gamma\cos2y}}\ ,\nonumber
\end{align}
where $\phi_{0}$ and $E$ are integration constants. Hence, we have to analyze
under which conditions on the parameter $\gamma$ and on the integration
constants $E$ and $\phi_{0}$ Eq. (\ref{caso2.22}) can have solutions
satisfying the conditions in Eq. (\ref{caso2.2}). As in the previous
subsection, it is possible to interpret Eq. (\ref{caso2.22}) as the energy
conservation of a one-dimensional problem in Eqs. (\ref{1Deffective1}) and
(\ref{1Deffective2}) in which $\phi_{1}$ plays the role of the effective time
$\tau=\phi_{1}$. However, in this case the boundary conditions in Eq.
(\ref{caso2.2}) become:%
\begin{equation}
A\left(  \tau+2\pi\right)  =A\left(  \tau\right)  \ .
\label{1Deffectiveboundcond2}%
\end{equation}
Thus, given two consecutive turning points $A_{0}$ and $A_{1}$:%
\[
A_{0}=\arccos\frac{\overline{E}}{2\gamma}\ ,\ \ \ A_{1}=2\pi-A_{0}\ ,
\]
one has to require that the the time needed to go from $A_{0}$ to $A_{1}$ is
half of the period in Eq. (\ref{1Deffectiveboundcond2}), namely:%
\[
\tau-\tau_{0}=\int_{A_{0}}^{A_{1}}\frac{dy}{\sqrt{2(\overline{E}-2\gamma\cos
y)}}=\pi\ .
\]
The time for a particle to go from $A_{0}$ to $A_{1}$ runs from $0$ to
infinity as $\overline{E}$ runs from $-2\gamma$ to $2\gamma$. This means that,
given a $\gamma$, there is always an $\overline{E}$ such that $\tau-\tau
_{0}=\pi$. Then, for $-2\gamma<$ $\overline{E}<2\gamma,$ it is always possible
to construct copies of the vacuum satisfying the strong boundary conditions.

Furthermore, it is easy to see that the norm of the copy is finite:%
\begin{align*}
\left\Vert U^{-1}dU\right\Vert  &  =\int_{T^{3}}d^{3}x\sqrt{g}tr\left(
U^{-1}dU\right)  ^{2}\\
&  =\left(  2\pi\right)  ^{2}\lambda_{1}\lambda_{2}\lambda_{3}\int_{T^{3}%
}d\phi_{1}\left(  \left(  \frac{\partial_{\phi_{1}}\alpha}{\lambda_{1}%
}\right)  ^{2}+\sin^{2}\alpha\left(  \left(  \frac{p}{\lambda_{2}}\right)
^{2}+\left(  \frac{q}{\lambda_{3}}\right)  ^{2}\right)  \right)  <\infty\ .
\end{align*}
On the other hand, it worth emphasizing that while there is a common agreement
on the importance of the strong boundary conditions, it is not clear yet
whether or not it is mandatory to require the finite norm condition. For
instance, it is possible to construct configurations of the Yang-Mills gauge
potential which have infinite norm but finite energy and/or action (see, for a
recent discussion, \cite{SorellaNorm}).

It is worth emphasizing that, at a first glance, one could think that the
presence of Gribov copies of the vacuum prevents one from using the Gribov
semi-classical approach whose aim is, of course, to eliminate Gribov copies
from a suitable neighborhood of the vacuum itself. However, the copies which
can be eliminated using the Gribov semiclassical approach are "small" copies,
namely zero-modes of the Faddeev-Popov operator (which is a linear condition).
On the other hand, the vacuum copies which have been constructed here are
solutions of the full non-linear equation for the copies and they would
disappear if one would only consider the linearized equation (which reduces
just to the Laplace equation in all the cases discussed in the present paper).
Therefore, the Gribov semiclassical approach can be applied, for instance, to
the $R\times T^{3}$ or $T^{4}$ cases. For instance, in the $T^{4}$ case in the
Landau gauge, the inverse Gribov propagator $\left(  \Delta_{G}\right)  ^{-1}$
would have, schematically, the usual form%
\[
\left(  \Delta_{G}\right)  ^{-1}=\square+\gamma\square^{-1}\ ,
\]
where $\gamma$ is the non-perturbative Gribov mass parameter determined by the
usual gap equation \cite{Gri78} and $\square$ is the Laplacian on $T^{4}$.
Hence, the only technical difference with respect the usual case would be
that, in order to invert the Laplacian one should use the Fourier series
instead of the Fourier transform. Indeed, in this way one can eliminate, from
a suitable neighborhood of the vacuum $A_{\mu}=0$, zero modes of the
Faddeev-Popov operator but the copies of the vacuum discussed above cannot be
eliminated with this procedure. The physical consequences of this fact are
very interesting also from the point of view of lattice QCD and are actually
under investigation (for the relations between the continuous analysis of
Gribov copies and the lattice see also \cite{new1}, \cite{new2} and
\cite{new3}).

\subsection{$%
\mathbb{R}
\times T^{2}$ topology}

As it has been already explained, this topology\textbf{ }is very important,
for instance, in relation with the infinite volume limit of Yang-Mills theory
as well as chiral symmetry breaking.

Actually, as far as the infinite volume limit and chiral symmetry breaking are
concerned (see, for instance, \cite{infinitevolume} \cite{casimir1}), the
setting corresponds to the Euclidean theory defined on $%
\mathbb{R}
\times T^{3}$. Thus, in this case the Landau gauge is to be preferred.
However, the present construction trivially extends to this situation as well.
One has just to interpret the coordinate $x$ in the metric in Eq.
(\ref{metr2.5}) as the Euclidean time in order to apply the present analysis
to this case.

Here we will consider the following ansatz corresponding to the metric in Eq.
(\ref{metr2.5}):
\begin{align}
\alpha &  =\alpha(x)\ ,\ \ \widehat{n}^{1}=\cos\left(  p\phi_{1}+q\phi
_{2}\right)  \ ,\ \ \widehat{n}^{2}=\sin\left(  p\phi_{1}+q\phi_{2}\right)
\ ,\label{caso3.1}\\
\widehat{n}^{3}  &  =0\ ,\ \ p,q\in%
\mathbb{Z}
\ .\nonumber
\end{align}
It is easy to see that the above ansatz satisfies the conditions in Eqs.
(\ref{hedgehogcondition3}) and (\ref{hedgehogcondition4}) with the following
$L$:%
\[
L=-\frac{1}{2}\left(  \left(  \frac{p}{\lambda_{2}}\right)  ^{2}+\left(
\frac{q}{\lambda_{3}}\right)  ^{2}\right)  \ .
\]
Thus, in this case the equation (\ref{sinegordon1.5}) for the copy of the
vacuum reduces to a flat elliptic sine-Gordon equation:%
\begin{align}
\left(  \frac{\partial^{2}}{\partial x^{2}}\right)  \alpha &  =\gamma
\sin2\alpha\ ,\label{caso3.2}\\
\gamma &  =\frac{\lambda_{1}^{2}}{2}\left(  \left(  \frac{p}{\lambda_{2}%
}\right)  ^{2}+\left(  \frac{q}{\lambda_{3}}\right)  ^{2}\right)  \ .
\label{caso3.3}%
\end{align}
In order for a solution of Eq. (\ref{caso3.2}) to define a copy of the vacuum
satisfying the strong boundary condition (see Eqs. (\ref{hedgehog1.75}),
(\ref{tub3.5}) and (\ref{tub3.75})) it is necessary to require%
\begin{equation}
\underset{x\rightarrow+\infty}{\lim}\alpha(x)=m\pi\ ,\ \ \underset
{x\rightarrow-\infty}{\lim}\alpha(x)=n\pi\ ,\ \ m,n\in%
\mathbb{Z}
\ . \label{caso3.4}%
\end{equation}
Similarly to the previous subsections, Eq. (\ref{caso3.2}) can be reduced to:%
\begin{align}
E  &  =\frac{1}{2}\left[  \left(  \frac{\partial\alpha}{\partial x}\right)
^{2}+\gamma\cos2\alpha\right]  \ ,\label{caso3.33}\\
x-x_{0}  &  =\pm\int_{\alpha(x_{0})}^{\alpha(x)}\frac{dy}{\sqrt{2E-\gamma
\cos2y}}\ ,\nonumber
\end{align}
where $x_{0}$ and $E$ are integration constants. Hence, we have to analyze
under which conditions on the parameter $\gamma$ and on the integration
constants $E$ and $x_{0}$ Eq. (\ref{caso3.33}) can have solutions satisfying
the conditions in Eq. (\ref{caso3.4}) for some $m$ and $n$. As in the previous
subsections, (\ref{caso3.33}) has the form of energy conservation of a
one-dimensional problem (\ref{1Deffective1}) with the identifications
(\ref{1Deffective2}). In this case the strong boundary conditions in Eq.
(\ref{caso3.4}) become:%
\begin{equation}
\underset{x\rightarrow+\infty}{\lim}A(x)=2m\pi\ ,\ \ \underset{x\rightarrow
-\infty}{\lim}A(x)=2n\pi\ ,\ \ m,n\in%
\mathbb{Z}
\ . \label{caso3.5}%
\end{equation}

For $m=1$ and $n=0$ this means that $0$ and $2\pi$ have to be two turning
points with infinite period. As in (\ref{caso1.4}), this condition is ensured
for $\overline{E}=2\gamma$. Furthermore, in this case as well it is easy to
see that the norm of the copy is finite:%
\[
\left\Vert U^{-1}dU\right\Vert =\int_{T^{2}\times%
\mathbb{R}
}\sqrt{g}d^{3}xtr\left(  U^{-1}dU\right)  ^{2}<\infty\ .
\]
where%
\[
\alpha\left(  x\right)  =2\arctan\left[  \exp\left(  x-\bar{x}_{0}\right)
\right]  \ ,
\]
($\bar{x}_{0}$ being an integration constant) is the solution of Eq.
(\ref{caso3.33}).

\section{Elliptic Gribov Pendulum}

In this section, we will analyze the elliptic generalization of the Gribov
pendulum equation. It is easy to see that if one would try a naive
generalization of the spherical Gribov pendulum \cite{Gri78} to elliptic
coordinates (which reduces to the usual case in the spherical limit) then the
system of equations for the Gribov copies of the vacuum do not reduce to a
single scalar equation as it happens in the spherical case and, consequently,
it would be extremely difficult to analyze the corresponding system of
equations and the corresponding boundary conditions. On the other hand, within
the present framework, one is lead to the two ansatzs in Eqs.
(\ref{prolansatz}) and (\ref{oblansatz}) which \textit{do reduce} the
equations for the Gribov copies of the vacuum to a \textit{single scalar
equation}: this allows to study the strong boundary conditions in the usual
way. To the best of authors knowledge, this is the first non-trivial elliptic
generalization of the Gribov pendulum.

This analysis is particularly important since it sheds light on how sensible
the Gribov phenomenon is with respect to deformation of the spherical
symmetry. Even if the usual Gribov pendulum equation\footnote{Actually, the
whole issue of gauge copies is usually analyzed only in unbounded region.} is
analyzed in the unbounded region $r\in\left[  0,\infty\right[  $ \cite{Gri78}
\cite{SS05}, one implicitly assumes that similar results hold in a spherical
bounded region. The reason is that, of course, experimentally gluons can live
only within baryons and glueballs which are bounded regions. Therefore, it
makes sense to study the arising of gauge copies of the vacuum in bounded
region as well. Indeed, the results that no copy of the vacuum appear in the
case of the usual Gribov pendulum also holds in the case of a bounded
spherical region: as it will be now discussed, the elliptic case is more complicated.

\subsection{Prolate Spheroid}

\bigskip The line element for of flat three-dimensional Euclidean space in
prolate spheroidal coordinates is given by%
\begin{equation}
ds^{2}=a^{2}\left(  \sinh^{2}\mu+\sin^{2}\nu\right)  \left(  d\mu^{2}+d\nu
^{2}\right)  +a^{2}\sinh^{2}\mu\sin^{2}\nu d\phi^{2}%
\end{equation}
For a prolate spheroidal bounded region the coordinate ranges are given by
$0\leq\mu<R$, $\nu\in\left[  0,\pi\right]  $, $\phi\in\lbrack0,2\pi)$. The
coordinate $\mu$ represents the elliptic radius since $\mu=const$ surfaces are
ellipses. Such $\mu=const$ ellipses have different eccentricities: large
eccentricities (namely, large deformations from spherical symmetries)
correspond to small $\mu$ while small eccentricities (namely, small deviations
from spherical symmetry) correspond to large $\mu$. Thus, in the above
coordinates system, if one wants to analyze the limit of "large deformations
from spherical symmetry" then one has to consider the small $\mu$ region. On
the other hand, if one wants to consider the almost spherical case, then the
large $\mu$ limit has to be considered. As it will be shown in the following,
the large deformation limit is the most interesting case.

The Laplacian in this coordinates is given by%
\[
\nabla^{i}\partial_{i}=\frac{1}{a^{2}\left(  \sinh^{2}\mu+\sin^{2}\nu\right)
}\left[  \partial_{\mu}^{2}+\partial_{\nu}^{2}+\coth\mu\partial_{\mu}+\cot
\nu\partial_{\nu}\right]  +\frac{1}{a^{2}\sinh^{2}\mu\sin^{2}\nu}%
\partial_{\phi}^{2}%
\]

Then, the following ansatz satisfies the conditions in Eqs.
(\ref{hedgehogcondition3}) and (\ref{hedgehogcondition4})%
\begin{equation}
\alpha=\alpha\left(  \mu,\nu\right)  \text{ \ },\text{ \ }n^{1}=\cos\phi\text{
\ },\text{ \ }n^{2}=\sin\phi\text{ \ },\text{ \ }n^{3}=0\ , \label{prolansatz}%
\end{equation}
with the following $L$:%
\[
L=-\frac{1}{2a^{2}\sinh^{2}\mu\sin^{2}\nu}\ .
\]

\subsubsection{The strong boundary conditions}

Due to th fact that $\mu$ represents the elliptic radius, the strong boundary
conditions in an unbounded region corresponds to the following conditions on
$\alpha$:%
\begin{align}
&  \alpha\left(  \mu,\nu\right)  \underset{\mu\rightarrow0}{\rightarrow}%
n\pi+f(\mu,\nu)\ ,\label{probo1}\\
&  \alpha\left(  \mu,\nu\right)  \underset{\mu\rightarrow\infty}{\rightarrow
}m\pi\ ,\ \ \ n,m\in%
\mathbb{Z}
\ . \label{probo2}%
\end{align}
where $f(\mu,\nu)$ ensures regularity of the copies. On the other hand, if one
is analyzing the theory in a bounded region of elliptic radius $R$ then the
condition in Eq. (\ref{probo2}) has to be replaced by%
\begin{equation}
\alpha\left(  \mu,\nu\right)  \underset{\mu\rightarrow R}{\rightarrow}%
m\pi\ ,\ \ \ m\in%
\mathbb{Z}
\ , \label{probo3}%
\end{equation}
since the boundary of the region is the surface $\mu=R$ and one has to require
that the gauge copy belongs to the center of the gauge group on the boundary
of the region itself.

\subsubsection{Regularity conditions}

In the prolate spheroidal case, the inverse of the Jacobian in the
transformation (\ref{regularity2}) reads
\begin{equation}
J^{-1}=\left(  \frac{\partial(\mu,\nu,\phi)}{\partial(x,y,z)}\right)
=\frac{1}{a}\left(
\begin{array}
[c]{ccc}%
\frac{\cosh\mu\sin\nu\cos\phi}{\sinh^{2}\mu+\sinh^{2}\nu} & \frac{\cosh\mu
\sin\nu\sin\phi}{\sinh^{2}\mu+\sinh^{2}\nu} & \frac{\sinh\mu\cos\nu}{\sinh
^{2}\mu+\sinh^{2}\nu}\\
\frac{\sinh\mu\cos\nu\cos\phi}{\sinh^{2}\mu+\sinh^{2}\nu} & \frac{\sinh\mu
\cos\nu\sin\phi}{\sinh^{2}\mu+\sinh^{2}\nu} & -\frac{\cosh\mu\sin\nu}%
{\sinh^{2}\mu+\sinh^{2}\nu}\\
-\frac{\sin\phi}{\sinh\mu\sin v} & \frac{\cos\phi}{\sinh\mu\sin\nu} & 0
\end{array}
\right)  \label{probo4}%
\end{equation}
Thus, singularities could appear for $\mu=0$ and $\nu=0$. Following the
prescription in Eq. (\ref{regularity4}), near points with $\mu=0$ and/or
$\nu=0$ copies behave as%
\begin{align*}
R_{\mu}^{c}  &  \rightarrow\beta\widehat{n}^{c}\partial_{\mu}f\\
R_{\nu}^{c}  &  \rightarrow\beta\widehat{n}^{c}\partial_{\nu}f\\
R_{\phi}^{c}  &  \rightarrow\beta^{2}\delta^{dc}f^{2}\varepsilon_{abd}%
\widehat{n}^{a}\partial_{\phi}\widehat{n}^{b}+\beta f\partial_{\phi}%
\widehat{n}^{c}%
\end{align*}
where $f$ must be such that functions (\ref{regularity5}) are regular. A
sufficient condition which ensures regularity is:%
\[
f(\mu,\nu)\underset{\mu,\nu\rightarrow0}{\rightarrow}\mu^{3}\nu^{3}%
\]
Besides the singularity at $\mu=0$, there are singularities for $\nu=0$ and
$\nu=\pi$ as well (which are similar to the $1/\sin\theta$ singularity in the
spherical case). However, in the elliptic case, the $\sin\nu$ factor is not
automatically canceled by the internal vector $\widehat{n}^{i}$ of the
hedgehog ansatz (which only depends on $\phi$ in the present case). Therefore
the profile function $\alpha(\mu,\nu)$ has to take care of this divergence.

\subsubsection{\bigskip Prolate pendulum}

Eq. (\ref{sinegordon1.5}) in the prolate elliptic coordinates reduces to the
following elliptic prolate Gribov pendulum:%
\begin{equation}
\frac{1}{\sinh^{2}\mu+\sin^{2}\nu}\left[  \partial_{\mu}^{2}+\partial_{\nu
}^{2}+\coth\mu\partial_{\mu}+\cot\nu\partial_{\nu}\right]  \alpha=\frac
{1}{2\sinh^{2}\mu\sin^{2}\nu}\sin2\alpha\ . \label{prolate equation}%
\end{equation}
Now we consider the limits $\mu\rightarrow0$ and $\mu\rightarrow\infty$ which
correspond to the cases of large and small deformation from spherical symmetry respectively.

\begin{itemize}
\item For $\mu\rightarrow0$, equation (\ref{prolate equation}) takes the form:%
\begin{equation}
\left[  \partial_{\mu}^{2}+\partial_{\nu}^{2}+\coth\mu\partial_{\mu}+\cot
\nu\partial_{\nu}\right]  \alpha=\frac{1}{2\sinh^{2}\mu}\sin2\alpha\ ,
\label{prolate2}%
\end{equation}
Interestingly enough, in this limit the ansatz $\alpha=\alpha\left(
\mu\right)  $ is consistent and the above equation reduces to%
\begin{equation}
\frac{d^{2}\alpha}{d\mu^{2}}+\coth\mu\frac{d\alpha}{d\mu}=\frac{1}{2\sinh
^{2}\mu}\sin2\alpha\ , \label{prolate3}%
\end{equation}
With the following change of coordinate%
\begin{align}
\tau &  =\ln\left\vert \tanh\left(  \frac{\mu}{2}\right)  \right\vert
\label{spaccrpolate}\\
\mu &  =2\operatorname{arcth}e^{\tau}\\
\sinh\left(  \operatorname{arcth}x\right)   &  =\frac{x}{\sqrt{1-x^{2}}}%
\end{align}
the equation (\ref{prolate3}) can be written as%
\[
\frac{d^{2}\alpha\left(  \tau\right)  }{d\tau^{2}}=\frac{1}{2}\sin
2\alpha\left(  \tau\right)  \ .
\]
Of course, in this case, the natural boundary conditions correspond to the
analysis \ within a bounded region in Eqs. (\ref{probo1}) and (\ref{probo3})
in which, in suitable units, the radius $R$ is very small%
\[
R\ll1.
\]
In terms of the coordinates $\tau$ in Eq. (\ref{spaccrpolate}) Eqs.
(\ref{probo1}) and (\ref{probo3})%
\[
\alpha\left(  \tau\right)  \underset{\tau\rightarrow-\infty}{\rightarrow}%
2n\pi\ ,
\]%
\[
\alpha\left(  \tau\right)  \underset{\tau\rightarrow\tau^{\ast}}{\rightarrow
}2m\pi\ ,\ \ \ n,m\in%
\mathbb{Z}
\ .
\]
In the limit of large prolate deformations, the equation can be integrated
analytically:%
\begin{align}
E  &  =\frac{1}{2}\left[  \left(  \frac{d\alpha}{d\tau}\right)  ^{2}+\frac
{1}{2}\cos2\alpha\right]  \ ,\label{profinal1}\\
\tau-\tau_{0}  &  =\pm\int_{\alpha(\tau_{0})}^{\alpha(\tau)}\frac{dy}%
{\sqrt{2E-\frac{1}{2}\cos2y}}\ . \label{profinal2}%
\end{align}
This result discloses in a very clear way how sensible the Gribov phenomenon
is not only to topology but also to the shapes of the region where it is
analyzed. Indeed, even if the present analytic solution cannot satisfy the
regularity conditions in $\nu=0$ and $\nu=\pi$ (which are the north and south
poles of the ellipse) since it is $\nu-$independent, a small deformation of
the ellipse at the poles could eliminate the necessity to require regularity
of the solution at $\nu=0$ and $\nu=\pi$ and consequently could give rise to
the sudden appearance of a copy of the vacuum. Hence, the present analysis
strongly suggests that large prolate deformations from spherical symmetry can
be quite pathological. These results can be relevant in the cases in which one
is analyzing Yang-Mills theory in bounded regions.
\end{itemize}

\bigskip

\begin{itemize}
\item For $\mu\rightarrow\infty,$ equation (\ref{prolate equation}) reduces
to:%
\[
\left[  \partial_{\mu}^{2}+\partial_{\nu}^{2}+\partial_{\mu}+\cot\nu
\partial_{\nu}\right]  \alpha=\frac{1}{2}\csc^{2}\nu\sin2\alpha\ .
\]
Since the large $\mu$ limit corresponds to small deviations from spherical
symmetry, in this case one should recover the standard results on the absence
of vacuum copies.
\end{itemize}

\subsection{Oblate Spheroid}

The line element for a flat three-dimensional Euclidean space in oblate
spheroidal coordinates is given by%
\[
ds^{2}=a^{2}\left(  \sinh^{2}\mu+\sin^{2}\nu\right)  \left(  d\mu^{2}+d\nu
^{2}\right)  +a^{2}\cosh^{2}\mu\cos^{2}\nu d\phi^{2}%
\]
For a oblate spheroidal bounded region the coordinate ranges are given by
$0\leq\mu<\mu^{\ast},\nu\in\left[  0,\pi\right]  ,\phi\in\lbrack0,2\pi)$. Also
in this case $\mu$ is the elliptic radius since $\mu=const$ surfaces are
ellipses with eccentricities which decrease with $\mu$. As in the prolate
case, if one wants to analyze the limit of "large deformations" then one has
to consider the small $\mu$ region. On the other hand, if one wants to
consider the almost spherical case, then the large $\mu$ limit has to be considered.

The Laplacian in this coordinates reads%
\[
\nabla^{i}\partial_{i}=\frac{1}{a^{2}\left(  \sinh^{2}\mu+\sin^{2}\nu\right)
}\left[  \partial_{\mu}^{2}+\partial_{\nu}^{2}+\tanh\mu\partial_{\mu}+\tan
\nu\partial_{\nu}\right]  +\frac{1}{a^{2}\cosh^{2}\mu\cos^{2}\nu}%
\partial_{\phi}^{2}%
\]

Thus, the following ansatz satisfies the conditions in Eqs.
(\ref{hedgehogcondition3}) and (\ref{hedgehogcondition4})%
\begin{equation}
\alpha=\alpha\left(  \mu,\nu\right)  \text{ \ },\text{ \ }n^{1}=\cos\phi\text{
\ },\text{ \ }n^{2}=\sin\phi\text{ \ },\text{ \ }n^{3}=0 \label{oblansatz}%
\end{equation}
with the following $L$:%
\[
L=-\frac{1}{2a^{2}\cosh^{2}\mu\cos\nu}\ .
\]

\subsubsection{Strong boundary conditions}

As in the prolate case, the strong boundary conditions in an unbounded region
corresponds to the following conditions on $\alpha$:%
\begin{align}
&  \alpha\left(  \mu,\nu\right)  \underset{\mu\rightarrow0}{\rightarrow}%
2n\pi\ ,\label{oblobo1}\\
&  \alpha\left(  \mu,\nu\right)  \underset{\mu\rightarrow\infty}{\rightarrow
}2m\pi\ ,\ \ \ n,m\in%
\mathbb{Z}
\ . \label{oblobo2}%
\end{align}
On the other hand, if one is analyzing the theory in a bounded region of
elliptic radius $R$ then the condition in Eq. (\ref{oblobo2}) has to be
replaced by%
\begin{equation}
\alpha\left(  \mu,\nu\right)  \underset{\mu\rightarrow R}{\rightarrow}%
2m\pi\ ,\ \ \ n,m\in%
\mathbb{Z}
\ . \label{oblobo3}%
\end{equation}

\paragraph{Regularity conditions}

As it has been explained in the previous subsection, the informations about
the singularities at the origin are encoded in the Jacobian. Also in this
case, besides the singularity for $\mu=0$, there are singularities for $\nu=0$
and $\nu=\pi$ as well Since the $\sin\nu$ factor is not automatically canceled
by the internal vector $\widehat{n}^{i}$ of the hedgehog ansatz, the profile
function $\alpha(\mu,\nu)$ has to take care of this divergence: a sufficient
condition to ensure regularity is:%
\[
f(\mu,\nu)\underset{\mu,\nu\rightarrow0}{\rightarrow}\mu^{3}\nu^{3}%
\]

\subsubsection{Oblate pendulum}

Hence, the equation for copies of the vacuum (\ref{sinegordon1.5}) in the
oblate case now takes the form of the following oblate Gribov pendulum%
\begin{equation}
\frac{1}{\sinh^{2}\mu+\sin^{2}\nu}\left[  \partial_{\mu}^{2}+\partial_{\nu
}^{2}+\tanh\mu\partial_{\mu}+\tan\nu\partial_{\nu}\right]  \alpha=\frac
{1}{2\cosh^{2}\mu\cos^{2}\nu}\sin\left(  2\alpha\right)  \ .
\label{oblate equation}%
\end{equation}
Now we consider the limits $\mu\rightarrow0$ and $\mu\rightarrow\infty$
corresponding to large and small deformation from spherical symmetry respectively.

\begin{itemize}
\item \bigskip For $\mu\rightarrow0$ (\ref{oblate equation}) reduces to%
\[
\left[  \partial_{\mu}^{2}+\partial_{\nu}^{2}+\tan\nu\partial_{\nu}\right]
\alpha=\frac{1}{2}\tan^{2}\nu\sin\left(  2\alpha\right)  \ .
\]
Unlike the prolate case, in the present case it is not possible to find
analytic solutions in the limit of large deformations.

\item For $\mu\rightarrow\infty$ (\ref{oblate equation}) reduces to%
\begin{equation}
\left[  \partial_{\mu}^{2}+\partial_{\nu}^{2}+\partial_{\mu}+\tan\nu
\partial_{\nu}\right]  \alpha=\frac{\tanh^{2}\mu}{2\cos^{2}\nu}\sin\left(
2\alpha\right)
\end{equation}

\end{itemize}

Also in this case, since the large $\mu$ limit corresponds to small deviations
from spherical symmetry, one should recover the standard results on the
absence of vacuum copies.\bigskip

It is worth emphasizing that in the large deformation limit the prolate and
oblate Gribov pendulum equations (\ref{prolate equation}) and
(\ref{oblate equation}) differ significantly. In particular, unlike the
spherical or oblate cases, in the prolate case the Gribov pendulum equation
can be integrated exactly. This strongly suggests that prolate deformations
from spherical symmetry are more pathological than oblate deformations.
Indeed, the present results strongly suggest that the strong deformation limit
of Eq. (\ref{prolate equation}) may support copies of the vacuum. This
analysis appears to be quite relevant as far as the issue of gauge copies in
bounded region is concerned.

\section{Conclusions and further comments}

In this paper the arising of Gribov copies in regions with non-trivial
topologies (such as closed tubes $S^{1}\times D^{2}$, or $%
\mathbb{R}
\times T^{3}$) but flat metric has been analyzed. The technical tool has been
a generalization of the hedgehog ansatz beyond spherical symmetry. Such a
generalization of the hedgehog ansatz is very interesting in itself since, in
the case of the non-linear sigma-model, it provides one with a geometrical
recipe to reduce the the field equations of the non-linear sigma model (which
are a \textit{system of coupled non-linear partial differential equations}) to
a single scalar non-linear partial differential equation even when there is no
spherical symmetry. This ansatz allows to construct many analytic examples of
Gribov copies of the vacuum. Moreover, the elliptic Gribov pendulum has also
been derived (to the best of authors knowledge, for the first time) both in
the prolate and oblate cases. Our results suggest that large prolate
deformations from spherical symmetry are likely to be more pathological than
the oblate deformations. The requirement of absence of Gribov copies of the
vacuum satisfying the strong boundary conditions implies geometrical
constraints on the topology, on the shapes and on the sizes of the regions
with non-trivial topologies (such as upper bounds on the deviations from
spherical symmetry or constraint on the shape of the donut $S^{1}\times D^{2}%
$). Moreover, we have shown that in the case of a flat metric \textit{but with
the topology of} $T^{3}$ it is possible to construct copies of the vacuum
satisfying the strong boundary conditions and with finite norm. The present
results are interesting in relations with the infinite volume limit of
Yang-Mills theory (which is related to the $%
\mathbb{R}
\times T^{3}$\ topology). Indeed, one of the main point of the present
analysis has been to show that when the $T^{3}$ (as well as $T^{4}$) topology
\textit{explodes} to $R^{3}$ (or to $R^{4}$) the vacuum copies disappear since
neither.$R^{3}$ nor $R^{4}$ support copies of the vacuum. Hence, as far as the
Gribov copies are concerned, such limit is not smooth and should be studied
more carefully. This analysis is also relevant in all the cases in which
gluons are confined in regions of finite sizes with non-trivial topologies
such as in the cases of knotted flux tubes, lattice QCD and so on. Due to the
close relation between Gribov ambiguity and confinement the issue of the
Gribov copies in bounded regions (both on flat and on curved space-times) is
very important and worth to be investigated.

\section{Acknowledgments}

We thank Silvio Sorella for useful comments. This work is partially supported
by FONDECYT grant 1120352, and by the \textquotedblleft Southern Theoretical
Physics Laboratory\textquotedblright\ ACT-91 grant from CONICYT. The Centro de
Estudios Cient\'{\i}ficos (CECs) is funded by the Chilean Government through
the Centers of Excellence Base Financing Program of CONICYT. F. C. is also
supported by Proyecto de Inserci\'{o}n CONICYT 79090034, and by the Agenzia
Spaziale Italiana (ASI).

\section*{Appendix: \bigskip Landau gauge}

In this Appendix we will present two applications of the present generalized
hedgehog ansatz which lead to a novel realization of spherical and elliptical
symmetries respectively. The new way to implement spherical and elliptical
symmetry corresponds to a configuration in which the internal vectors $n^{i}$
of the generalized hedgehog ansatz depend only on the Euclidean time. This
realization of the spherical symmetry is only possible within the present
"generalized hedgehog" framework and it could be useful in the context of
calorons which are instantons which are periodic in Euclidean time. In both
cases we will construct the Landau gauge pendulum.

\subsection*{\bigskip A novel spherical Case}

Let us consider the following line element corresponding to a four-dimensional
euclidean space%
\[
ds^{2}=d\tau^{2}+dr^{2}+r^{2}\left(  d\theta^{2}+\sin^{2}\theta d\phi
^{2}\right)  \ ,\ \ \ 0\leq\tau\leq2\pi\ ,
\]
where $\tau$ plays the role of euclidean time (which is a periodic coordinate
of period $2\pi$) and the spatial section is written in spherical coordinates.
This situation is relevant in the cases in which one wants to describe finite
temperature effects. An ansatz satisfying conditions (\ref{hedgehogcondition3}%
) and (\ref{hedgehogcondition4}) suited to deal this situation is%
\[
\alpha=\alpha\left(  r\right)  \text{ \ },\text{ \ }n^{1}=\cos\left(
\omega\tau\right)  \text{ \ },\text{ \ }n^{2}=\sin\left(  \omega\tau\right)
\text{ \ },\text{ \ }n^{3}=0
\]
with
\[
L=-\frac{\omega^{2}}{2}%
\]

Then, the equation for Gribov copies (\ref{sinegordon1.5}) takes the form%
\begin{equation}
\partial_{r}^{2}\alpha+\frac{2}{r}\partial_{r}\alpha=\frac{\omega^{2}}{2}%
\sin\left(  2\alpha\right)  \ , \label{novelspherical}%
\end{equation}
which obviously \textit{is not equivalent} to the usual spherically symmetric
Gribov pendulum \cite{Gri78} \cite{SS05}. Defining $x=\ln r$ we can write the
above equation as%
\[
\partial_{x}^{2}\alpha+\partial_{x}\alpha=\frac{\omega^{2}}{2}e^{2x}%
\sin\left(  2\alpha\right)  \ ,
\]
another useful form is, defining $y=-1/r$, the following
\[
\partial_{y}^{2}\alpha=\frac{\omega^{2}}{2y^{4}}\sin\left(  2\alpha\right)
\]
The regularity condition at the origin can be analyzed as in the previous sections.

\subsection*{Prolate Spheroidal Case}

\bigskip Let us consider the line element of four-dimensional Euclidean
space-time (in which the Euclidean time has been compactified to describe
finite-temperature effects) in prolate elliptic coordinates%
\[
ds^{2}=d\tau^{2}+a^{2}\left(  \sinh^{2}\mu+\sin^{2}\nu\right)  \left(
d\mu^{2}+d\nu^{2}\right)  +a^{2}\sinh^{2}\mu\sin^{2}\nu d\phi^{2}%
\]
An ansatz satisfying conditions (\ref{hedgehogcondition3}) and
(\ref{hedgehogcondition4}) suited to deal this situation is%
\[
\alpha=\alpha\left(  \mu,\nu\right)  \text{ \ },\text{ \ }n^{1}=\cos\left(
\omega\tau\right)  \text{ \ },\text{ \ }n^{2}=\sin\left(  \omega\tau\right)
\text{ \ },\text{ \ }n^{3}=0
\]
with
\[
L=-\frac{\omega^{2}}{2}%
\]

Then, Eq. (\ref{sinegordon1.5}) reduces to%
\[
\frac{1}{a^{2}\left(  \sinh^{2}\mu+\sin^{2}\nu\right)  }\left[  \partial_{\mu
}^{2}+\partial_{\nu}^{2}+\tanh\mu\partial_{\mu}+\tan\nu\partial_{\nu}\right]
\alpha=\frac{\omega^{2}}{2}\sin\left(  2\alpha\right)
\]
Also in this case, the regularity conditions can be studied as in the previous sections.

\end{document}